\documentclass{pasj00}

\begin{document}
\SetRunningHead{Takita et al.}{Calibration of AFASS data.}

\title{Calibration of the AKARI Far-infrared All Sky Survey Maps.}


%
\author{%
  Satoshi   \textsc{Takita},    \altaffilmark{1}
  Yasuo     \textsc{Doi},       \altaffilmark{2}
  Takafumi  \textsc{Ootsubo},   \altaffilmark{2}
  Ko        \textsc{Arimatsu},  \altaffilmark{1,3}
  Norio     \textsc{Ikeda},     \altaffilmark{1}
  Mitsunobu \textsc{Kawada},    \altaffilmark{1}
  Yoshimi   \textsc{Kitamura},  \altaffilmark{1}
  Shuji     \textsc{Matsuura},  \altaffilmark{1}
  Takao     \textsc{Nakagawa},  \altaffilmark{1}
  Makoto    \textsc{Hattori},   \altaffilmark{4}
  Takahiro  \textsc{Morishima}, \altaffilmark{4}
  Masahiro  \textsc{Tanaka},    \altaffilmark{5}
  and
  Shinya    \textsc{Komugi},    \altaffilmark{6}
}
\altaffiltext{1}{Institute of Space and Astronautical Science, Japan Aerospace Exploration Agency,
  3-1-1 Yoshinodai, Chuo, Sagamihara, Kanagawa 252-5210, Japan}
\email{takita@ir.isas.jaxa.jp}
\altaffiltext{2}{Department of Earth Science and Astronomy, Graduate School of Arts and Sciences, The University of Tokyo,
  Komaba 3-8-1, Meguro, Tokyo 153-8902, Japan}
\altaffiltext{3}{Department of Astronomy, Graduate school of Science, The University of Tokyo,
  7-3-1 Hongo, Bunkyo, Tokyo 113-0033, Japan}
\altaffiltext{4}{Astronomical Institute, Graduate School of Science, Tohoku University,
  Aramaki, Aoba-ku, Sendai 980-8578, Japan}
\altaffiltext{5}{Center for Computational Sciences, Tsukuba University,
  1-1-1 Tennodai, Tsukuba-city, Ibaraki 305-8577, Japan}
\altaffiltext{6}{National Astronomical Observatory of Japan,
  2-21-1 Osawa, Mitaka, Tokyo, Japan}


\maketitle

\begin{abstract}
We present an initial analysis of the properties of the all-sky image obtained by the Far-Infrared Surveyor (FIS) onboard the AKARI satellite, at 65~$\mu$m (N60), 90~$\mu$m (WIDE-S), 140~$\mu$m (WIDE-L),and 160~$\mu$m (N160).
Absolute flux calibration was determined by comparing the data with the COBE/DIRBE data sets, and the intensity range was as wide as from a few MJy~sr$^{-1}$ to $>$1~GJy~sr$^{-1}$.
The uncertainties are considered to be the standard deviations with respect to the DIRBE data, and they are less than 10\% for intensities above 10, 3, 25, and 26~MJy~sr$^{-1}$ at the N60, WIDE-S, WIDE-L, and N160 bands, respectively.
The characteristics of point sources in the image were also determined by stacking maps centred on photometric standard stars.
The full width at half maxima of the point spread functions (PSFs) were 63$''$, 78$''$, and 88$''$ at the N60, WIDE-S, and WIDE-L bands, respectively.
The PSF at the N160 band was not obtained due to the sensitivity, but it is thought to be the same as that of the WIDE-L one.
\end{abstract}

\section{Introduction}
The Japanese infrared astronomical satellite AKARI \citep{murakami2007} performed an all sky survey at six infrared bands centred at 9, 18, 65, 90, 140 and 160~$\mu$m.
The four longer wavelength bands were observed with the Far-Infrared Surveyor (FIS), one of the two focal plane instruments onboard \citep{kawada2007}.
The two shorter FIS bands (N60 centred at 65~$\mu$m and WIDE-S centred at 90~$\mu$m) were observed by the SW detector, and the other two bands (WIDE-L at 140~$\mu$m and N160 at 160~$\mu$m) by the LW detector.
The all sky survey was started in 2006 May and covered $\sim$98~\% of the whole sky during its cold phase \citep{nakagawa2007}, continuing until helium exhaustion in 2007 Aug.

The All-sky Bright Source Catalogue (BSC, version 1), one of the primary goals of AKARI, was released in 2010 Mar \citep{fis_bsc}.
Although an all-sky survey image was not one of the primary goals of AKARI, it is scientifically valuable because extended sources are not covered by the BSC.
There are two previous All-sky Far-infrared (FIR) surveys that were obtained by the IRAS \citep{neugebauer1984} and COBE \citep{boggess1992} satellites.
The IRAS survey was made with two FIR photometric bands centred at 60 and 100~$\mu$m, with a spatial resolution of $\sim4'$.
The COBE mission expanded the wavelength coverage to 240~$\mu$m by using the DIRBE instrument \citep{hauser1998}, but the spatial resolutions were only $\sim$0\fdg7.
Recently, \citet{iris} created a new data set of IRIS (Improved Reprocessing of the IRAS Survey) by combining the IRAS and COBE data.
The IRIS is referred in many fields, but it only covers the wavelengths at or shorter than 100~$\mu$m.
Since the thermal radiation from interstellar dust, whose typical temperature is $\sim$30~K, has its peak at 100--200~$\mu$m, the longer wavelength data are important.
The present AKARI all-sky image includes data at 65, 90, 140, and 160~$\mu$m, with spatial resolution of $\sim$1\farcm5.
Thus we have created an all sky survey map from the AKARI data set \citep{doi2009} to provide new legacy data which has a wide potential application for use by the scientific community in many studies.
In this paper, we report the calibration processes and performance of the FIS all sky maps.

\section{Data}

In the all sky survey mode, AKARI scanned the sky in an ecliptic orbit, with the telescope always pointing away from the earth.
At an altitude of $\sim$700~km, AKARI made the orbit around the earth in $\sim$100 minutes, scanning at a constant speed of $\sim$3\farcm6~s$^{-1}$.
We refer to the all sky image obtained in this way as the AFASS (AKARI Far-infrared All Sky Survey) image.

The observing strategy is explained in \citet{kawada2007}.
The detector was read out continuously with a constant sampling rate for each array, and reset to discharge the photocurrent at intervals of about 2 seconds nominally, or 0.5 seconds toward bright sky.
Close to the Galactic plane, where these long reset intervals were expected to lead to detector saturation, the detectors were reset for each sampling (correlated double sampling: CDS), with reset intervals corresponding to about 26 and 45 ms for the SW and LW detectors, respectively.
In this paper we refer to the ``Normal mode'' for the data which were taken with long reset intervals (i.e., 2 or 0.5 seconds) to distinguish it from the ``CDS mode'' for the short reset intervals.

The data were pre-processed using the AKARI pipeline tool called ``Green-Box''.
This pipeline applied basic calibrations such as corrections of non-linearity and sensitivity drifts of the detector, rejection of anomalous data due to charged particle hit (glitch), saturation, and other instrumental effects as well as dark-current subtraction.
Following this initial processing, another ``image'' pipeline optimised for imaging large spatial scale structures was applied.
At first, this image pipeline applied a transient response correction to the detector signal and subtraction of zodiacal emission.
Since the detector had a slow response, the detector output drifted along the scan causing some stripe patterns in the image.
To remove this artificial stripe pattern, we applied destriping.
This destriping procedure is as follows.
Firstly, we applied a two-dimensional Fourier transformation of the image.
Secondly, we masked the regions corresponding to the stripe patterns along the scan direction in wave number space.
Next, we interpolated the intensities from the surrounding regions into the masked regions.
Finally, we applied an inverse Fourier transformation to obtain the destriped image.
After the image destriping, we performed recursive deglitching and flat-fielding to produce the final processed image.
Details of the image pipeline are presented in \citet{doi2009, doi}.
The final image is made with a pixel scale of 15$''$.

\section{Calibrating the FIS Images}
\subsection{Calibration Method}

The FIS had a cold shutter whose temperature was $\sim$2.5~K and an internal calibration lamp, which had been well-calibrated in the laboratory.
However, the characteristics changed in-flight because of differences of the operation mode, surrounding environment, and modified sensitivities due to charged particle hits.
Thus, we needed to re-calibrate the FIS data using in-flight data.
In the FIS BSC, the flux densities of point sources were calibrated by measuring the photometric standard stars, the asteroids, and the planets Uranus and Neptune with well-known flux densities.
This calibration was made for point sources and the data reduction processes were highly optimized for extracting point sources, for which the optimal method differed from that needed to reduce the all-sky image (which included diffuse emission).
In addition to the normal all sky survey observations, AKARI also had capability to carry out pointed observations toward selected objects/regions with improved sensitivity.
In these pointed observations, the FIS intensity scales were determined on the basis of the COBE/DIRBE data by the self-flat-fielding method \citep{matsuura2007, ikeda2012}.
The calibration factors were obtained at those regions where the emission from the zodiacal light and the infrared cirrus dominated for the SW and LW detectors, respectively.
However, this method was applied only to the Normal mode data and the brightness range of the selected regions is of the order of 1--10~MJy~sr$^{-1}$, while the brightness reaches as high as $>$1~GJy~sr$^{-1}$ at the Galactic plane in the all sky survey.
Therefore, we needed to extend this method to include brighter emission and also the CDS mode data to be able to calibrate the AFASS image.

First of all, we estimated the expected brightnesses in the four FIS bands.
We used the DIRBE Zodi-Subtracted Mission Average (ZSMA) data set as an absolute brightness reference because it was one of the most reliable, well-calibrated all-sky data sets at far-infrared wavelengths to date.
The uncertainties in the absolute brightness were 10--14~\% for the 60--140~$\mu$m bands \citep{hauser1998}.
Although there was another data set of IRIS, it was also calibrated with the DIRBE, and it does not have $>$100~$\mu$m bands.
The expected brightnesses were calculated as follows:
First, we created continuous Spectral Energy Distributions (SEDs) by interpolating the DIRBE data with a power-low function, for each DIRBE pixel.
Since the DIRBE data were calibrated under the flat spectrum assumption, we made a colour correction using the applied power-law spectra.
Second, we multiplied the interpolated SEDs by the FIS relative spectral response function.
Third, we again made a colour correction to acquire the expected FIS values under the flat spectrum assumption.
Furthermore, since we selected ZSMA data, we calculated the brightnesses of the zodiacal emission in the FIS bands at the corresponding epoch using the Gorjian model \citep{gorjian2000}.
Here, we note that we only considered the ``smooth cloud'' component because remaining small-scale structures, such as asteroidal dust bands and a mean motion resonance were not well understood so far \citep{ootsubo}.
The sum of the ``DIRBE-expected'' and zodiacal emission model brightnesses should be compared to the observed FIS brightness.

\begin{figure*}
\begin{center}
\includegraphics[width=16cm]{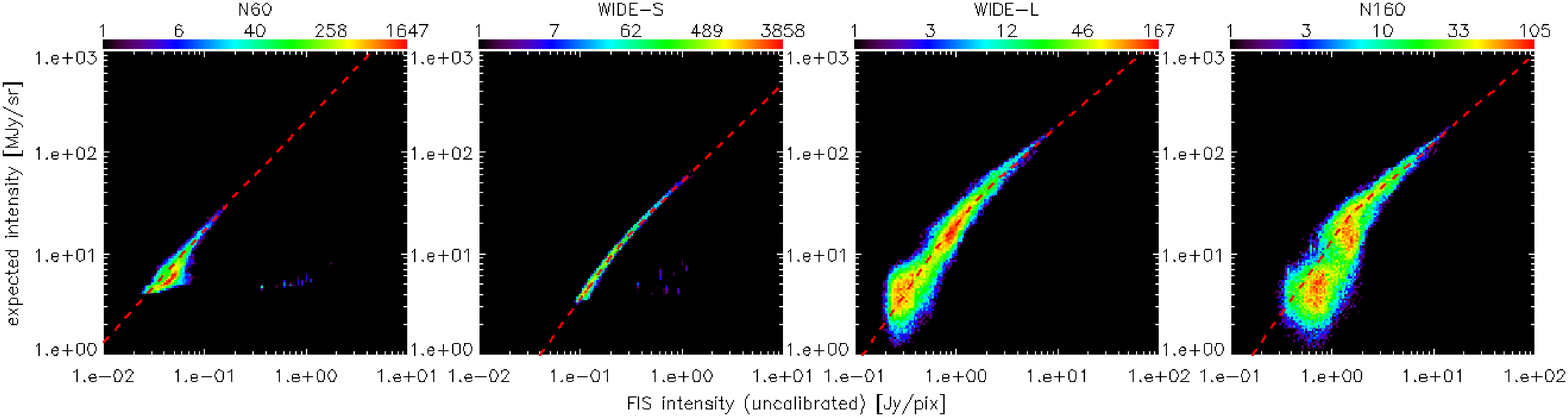}
\includegraphics[width=16cm]{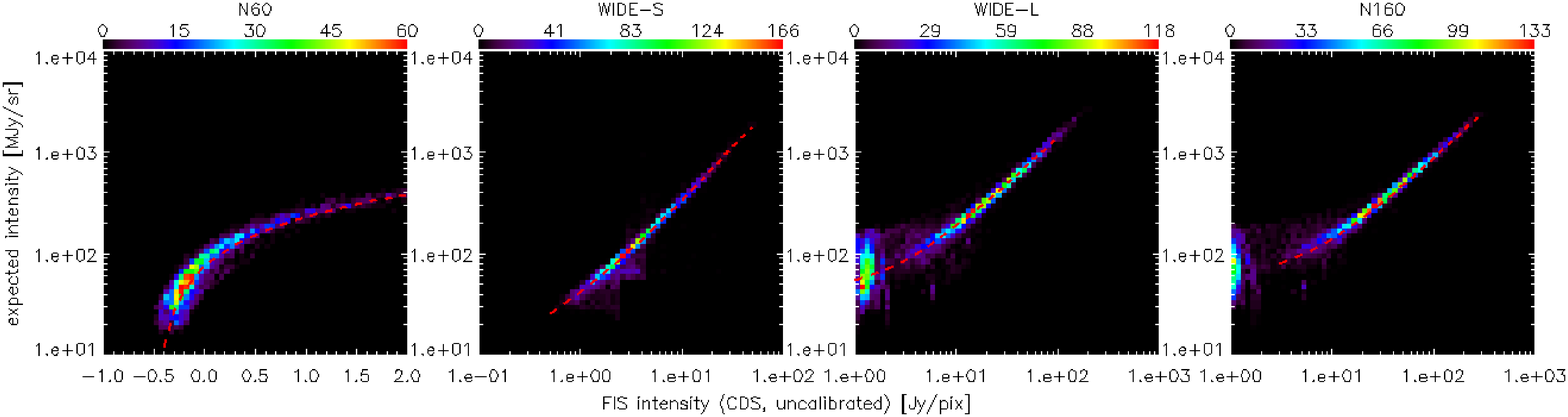}
\end{center}
\caption{
Relations between the Green-Box-precessed FIS intensity and the expected one based on the DIRBE data, including the zodiacal light emission.
The colour scales are the number of the DIRBE pixels, where $|\beta|>$40~degrees.
The top row panels are for the Normal mode data, and the bottom rows for the CDS mode.
The red broken lines show the adopted conversion factors.
}
\label{fig:cf}
\end{figure*}

To calibrate the FIS data, we also prepared the AFASS image.
The Green-Box tool converted the detector outputs to intensities in units of Jy~pixel$^{-1}$ with the conversion factors obtained in the laboratory.
As noted above, they were not well calibrated for the in-flight environment.
Although the image pipeline basically subtracted the zodiacal emission, we skipped this process to create image at this time.
Furthermore, since the intensity of the zodiacal light varies with time, we needed to restrict the epoch.
The all sky survey was the top priority during the first half year of AKARI's cold operations, with only a limited number of pointed observations made.
During the remaining time, the all sky survey was completed at high priority, although the fraction of time devoted to pointed observations was increased.
Thus, we chose data from the first half year (i.e., 2006 May to Oct), where there were few observational gaps due to pointed observations.
We convolved the FIS image with the beam pattern of the DIRBE at each DIRBE pixel to directly compare the FIS and DIRBE-expected brightnesses.
From this comparison, we estimated the conversion factors from the Green-Box-processed data into the absolute intensity.
As noted before, since we did not consider small-scale structures of the zodiacal emission, we excluded those regions of $|\beta$ (Ecliptic latitude)$|<$40~deg to avoid the effects of zodiacal emission.
The expected brightnesses of small-scale structures near the ecliptic plane are $<$5~MJy~sr$^{-1}$, $<$4~MJy~sr$^{-1}$, $<$1~MJy~sr$^{-1}$, and $<$1~MJy~sr$^{-1}$ for the N60, WIDE-S, WIDE-L, and N160 bands, respectively \citep{gorjian2000, ootsubo}.
After determining the conversion factors, we recreated the AFASS image using the conversion factors, and then compared with those predicted from the DIRBE data again.
This process was repeated several times, because the image pipeline does not guarantee the linear relation between the Green-Box calibrated FIS data and the processed image.
Figure~\ref{fig:cf} shows the converged conversion factors.

\subsection{Flux Uncertainties}
\begin{figure*}
\begin{center}
\includegraphics[width=16cm]{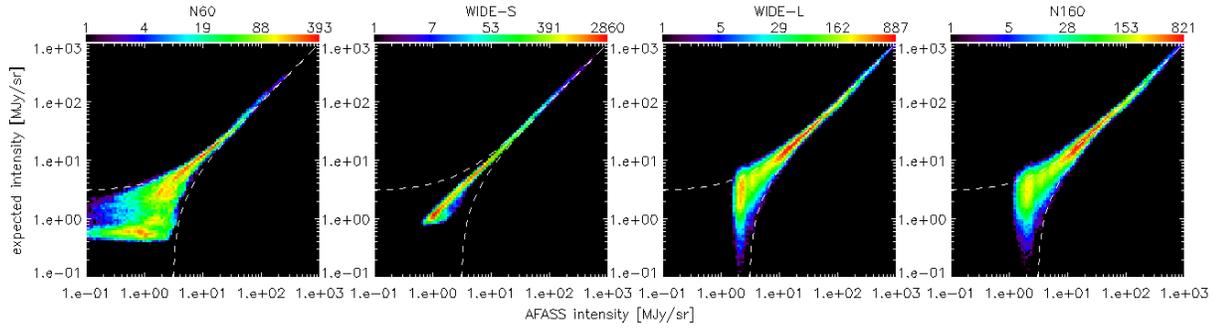}
\end{center}
\caption{
Relations between the calibrated FIS intensity and the expected intensity based on the DIRBE data, after subtraction of the smooth cloud component.
The colour scales are the number of the DIRBE pixels, where $|\beta|>$40~degrees.
The broken lines indicate the deviations of $\pm$3~MJy~sr$^{-1}$ from the linear relation of (expected intensity) = (calibrated intensity).
}
\label{fig:cal_result}
\end{figure*}

Figure~\ref{fig:cal_result} shows the pixel-to-pixel comparison between the calibrated FIS intensities and the expected values calculated from the DIRBE data, for each DIRBE pixel where $|\beta|>$40~deg.
In this figure, the zodiacal emission of the smooth cloud component was subtracted, and all-season data were used for the AFASS image.
We note that the FIS data still included other zodiacal emission components of asteroidal dust bands and a mean motion resonance.
A good correlation exists for brighter ($\sim$10~MJy~sr$^{-1}$) region, while there are some deviations from the linear relation at fainter brightness levels due to the incompleteness of the zodiacal emission model and the DIRBE sensitivity of $\gtrsim$10~MJy~sr$^{-1}$.
The intensity uncertainties were considered to be the standard deviations with respect to the DIRBE-expected intensities in Figure~\ref{fig:cal_result} for various intensity ranges, as shown in Table~\ref{tbl:cal_unc}.
We concluded that the standard deviations are less than 10~\% for $>$10, 3, 25, and 26~MJy~sr$^{-1}$ at the N60, WIDE-S, WIDE-L, and N160 bands, respectively.
We note that these estimates are based on the spatial resolution of DIRBE, i.e., about 0.7~$\deg$.
The WIDE-S band is the most sensitive one, and the two longer wavelength bands have large scatter at fainter (a few MJy~sr$^{-1}$) regions.
This is because the performance of the LW detector is not as good as that of the SW one, which can be also seen in Figure~\ref{fig:cf}.
Since the LW detector consisted of independent Ge:Ga elements, the sensitivity between the elements was largely scattered, while the SW detector was a monolithic Ge:Ga array and the sensitivity is quite uniform.
It is hard to estimate the sensitivity for the N60 band, because it is the one most seriously affected by the zodiacal emission of the 4 FIS bands.
Although we excluded the regions of $|\beta|<$40~deg, where the zodiacal emission is dominant, there still remain the small-scale structures of the emission at $|\beta|>$40~deg.
The faint intensity ($\lesssim$1~MJy~sr$^{-1}$) regions are thought to be affected by the zodiacal emission.
Therefore, we should re-calibrate the image when the model of the zodiacal emission becomes more reliable.

\begin{table}
\begin{center}
\caption{
Uncertainties in the intensity calibration for each band.
The uncertainties from the slope of unity in Figure~\ref{fig:cal_result} are given for individual intensity bins.
The 1~$\sigma$ uncertainties is calculated for each DIRBE pixel where $|\beta|>$40~deg.
}
\label{tbl:cal_unc}
\begin{tabular}{rrrrr}
\hline
& N60 & WIDE-S & WIDE-L & N160\\
\hline
MJy sr$^{-1}$\\
2.5 & $\ast$ & 15.1\% & $\ast$ & $\ast$\\
4.0 & $\ast$ &  8.1\% & $\ast$ & $\ast$\\
6.3 & 20.1\% &  4.7\% & $\ast$ & $\ast$\\
10  &  9.9\% &  4.5\% & 45.4\% & 56.1\%\\
16  &  8.4\% &  4.2\% & 14.3\% & 16.6\%\\
25  &  9.7\% &  4.3\% & 10.1\% & 11.2\%\\
40  & 10.8\% &  4.5\% &  8.3\% &  8.8\%\\
63  & 11.1\% &  5.3\% &  8.1\% &  7.9\%\\
100 & 12.5\% &  5.7\% &  8.9\% &  8.9\%\\
160 & 10.4\% &  7.3\% &  9.6\% &  9.7\%\\
250 & 13.2\% &  8.3\% &  9.8\% & 10.3\%\\
400 & 11.3\% &  8.1\% &  8.8\% &  8.9\%\\
630 & 11.7\% & 10.3\% &  7.0\% &  7.3\%\\
1000& 13.0\% &  7.9\% &  5.9\% &  6.5\%\\
\hline
\end{tabular}
\end{center}
\end{table}

\section{Point Sources}
\subsection{Point Spread Function}
\begin{figure*}
\begin{center}
\includegraphics[width=16cm]{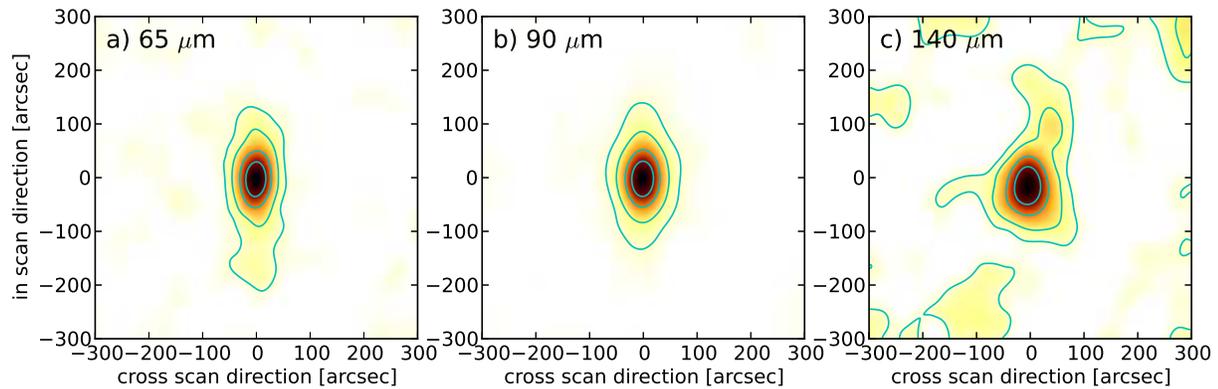}
\end{center}
\caption{
Obtained PSFs for the FIS N60 (65~$\mu$m), WIDE-S (90~$\mu$m), and WIDE-L (140~$\mu$m) bands.
The contours are at 75~\%, 50~\%, 25~\%, and 10~\% of the peak intensities.
}
\label{fig:map_psf}
\end{figure*}

\citet{arimatsu2014} have investigated the characteristics of point sources in the FIS maps using preliminary data, which were almost the same as the released data.
Here, we re-estimated the Point Spread Functions (PSFs) using the released data, following \citet{arimatsu2014}.
We chose 352 photometric standard stars from \citet{cohen1999} with expected flux densities of 0.02--10~Jy at the WIDE-S band.
To derive the PSF, we picked up those sources whose flux densities were greater than the 2.5~$\sigma$ detection limit to secure sufficient number of sources.
We then stacked 49, 97, and 9 stars with flux densities above 1.2, 0.28, and 0.7~Jy at the N60, WIDE-S, and WIDE-L bands, respectively.
The derived PSFs are shown in Figure~\ref{fig:map_psf} and their full width at half maximum (FWHM) are listed in Table~\ref{tbl:psf}.
We did not determine the PSF at the N160 band, because there were not enough bright standard stars detected at the band.
However, since the FIS PSFs are determined by the detector characteristics rather than optics and the PSF shapes at the WIDE-L and N160 bands are similar in pointed observations \citep{shirahata2009}, we assume that the N160 PSF is same as the WIDE-L one in the AFASS image.

\begin{table}
\begin{center}
\caption{FWHMs of stacked PSFs.}
\label{tbl:psf}
\begin{tabular}{crrr}
\hline
& N60 & WIDE-S & WIDE-L\\
\hline
FWHM [arcsec] & $63.4 \pm 0.2$ &  $77.8 \pm 0.2$ & $88.3 \pm 0.9$\\
(in-scan)     & $76.1 \pm 0.4$ & $102.3 \pm 0.3$ & $98.3 \pm 1.4$\\
(cross-scan)  & $31.3 \pm 0.3$ &  $55.0 \pm 0.1$ & $72.1 \pm 1.1$\\
solid angle [arcsec$^2$] & $2699 \pm 40$ & $6375 \pm 30$ & $8031 \pm 237$\\
\hline
N$_{\rm source}$ & 49 & 97 & 9\\
\hline
\end{tabular}
\end{center}
\end{table}

\subsection{Flux Calibrations}
\begin{figure}
\begin{center}
\includegraphics[width=8cm]{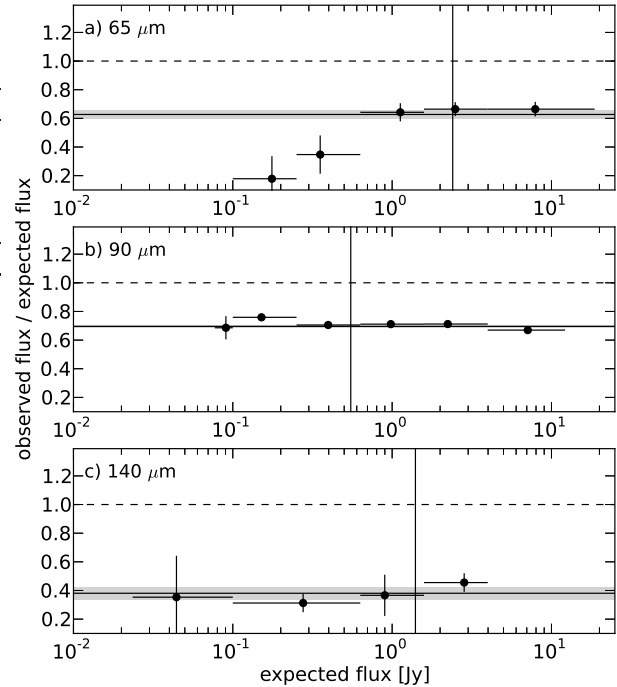}
\end{center}
\caption{
Observed-to-expected flux density ratio as a function of the expected flux density for the FIS (a) N60, (b) WIDE-S, and (c) WIDE-L bands.
The error bars on the y-axis represent the 1~$\sigma$ photometry uncertainties.
The solid lines with the filled areas are the weighted averages with 1~$\sigma$ uncertainties.
The dashed lines indicate the ratio of unity.
The vertical lines represent the 5~$\sigma$ detection limits of 2.4, 0.55, and 1.4~Jy for the All-Sky Survey at the FIS N60, WIDE-S, and WIDE-L bands, respectively \citep{kawada2007}.
}
\label{fig:map_pscal}
\end{figure}

We examined the flux calibrations for the point sources in the AFASS map.
We prepared $10'\times10'$ maps centred at the positions of selected standard stars.
We subtracted the sky background from each map by taking the median value, and normalized it to the expected flux density.
We separated selected sources into several groups over their appropriate flux density.
Then, we stacked the maps for each group and took average values.
We performed aperture photometry on the stacked maps using an aperture of 90$''$ radius with a sky background area of 120--300$''$ radius.
Figure~\ref{fig:map_pscal} shows the observed-to-expected flux density ratio for the stacked sources as a function of the expected flux density in each bin, and the mean ratios are 0.627$\pm$0.029, 0.696$\pm$0.008, and 0.381$\pm$0.043 for the N60, WIDE-S and WIDE-L bands, respectively.
Due to several improvements of the data reduction processes, these ratios differe from those of \citet{arimatsu2014}, which used preliminary data.
One significant difference is that the ratio at the N60 band becomes smaller for the faint source bins ($<$0.5~Jy).
However, these small ratios are likely caused by the poor sensitivity of the N60 band, i.e., the signal-to-noise ratios of the faint sources in each map are smaller than one.
Details of these processes are described in \citet{arimatsu2014}.

\subsection{Comparison with the FIS Bright Source Catalogue}
\begin{figure}
\begin{center}
\includegraphics[width=8cm]{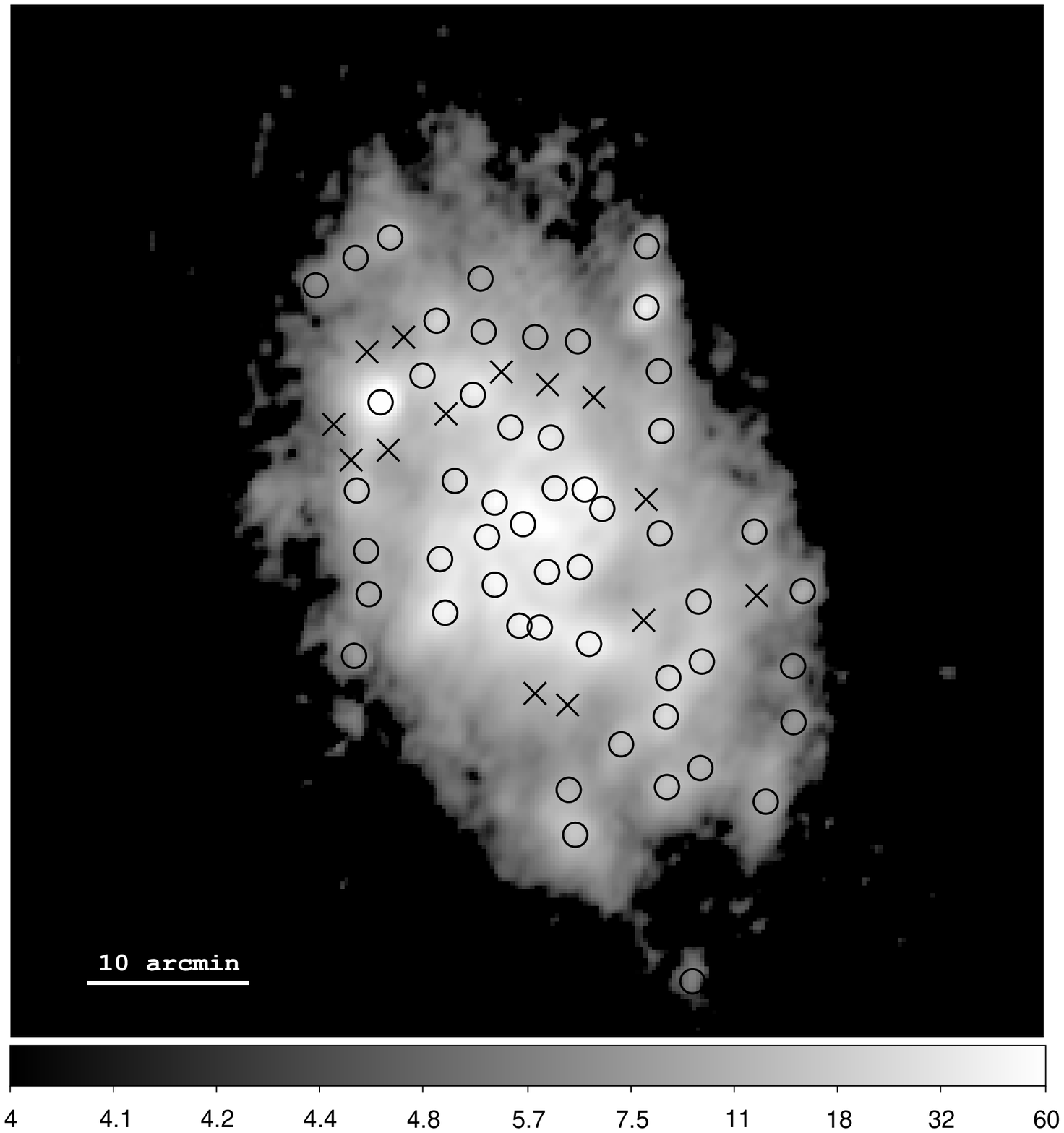}
\end{center}
\caption{
Example of the centroid determination of BSC sources around M33 on the AFASS map in the WIDE-S band.
The circle symbols are the sources whose centroids were determined, and the cross ones not well determined.
}
\label{fig:map_ex}
\end{figure}

\begin{figure*}
\begin{center}
\includegraphics[width=16cm]{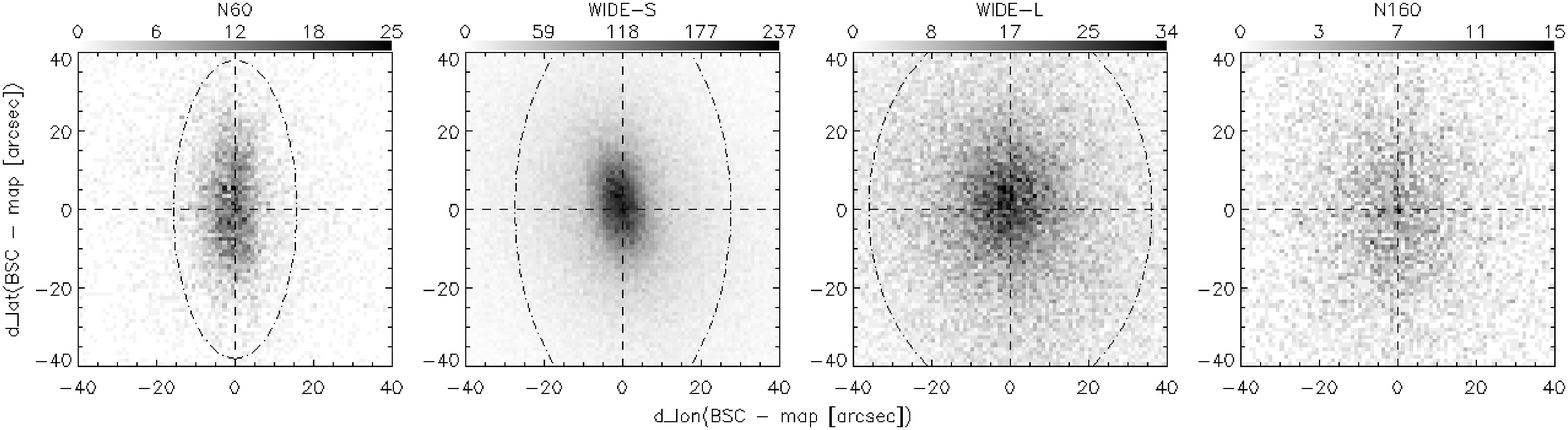}
\includegraphics[width=16cm]{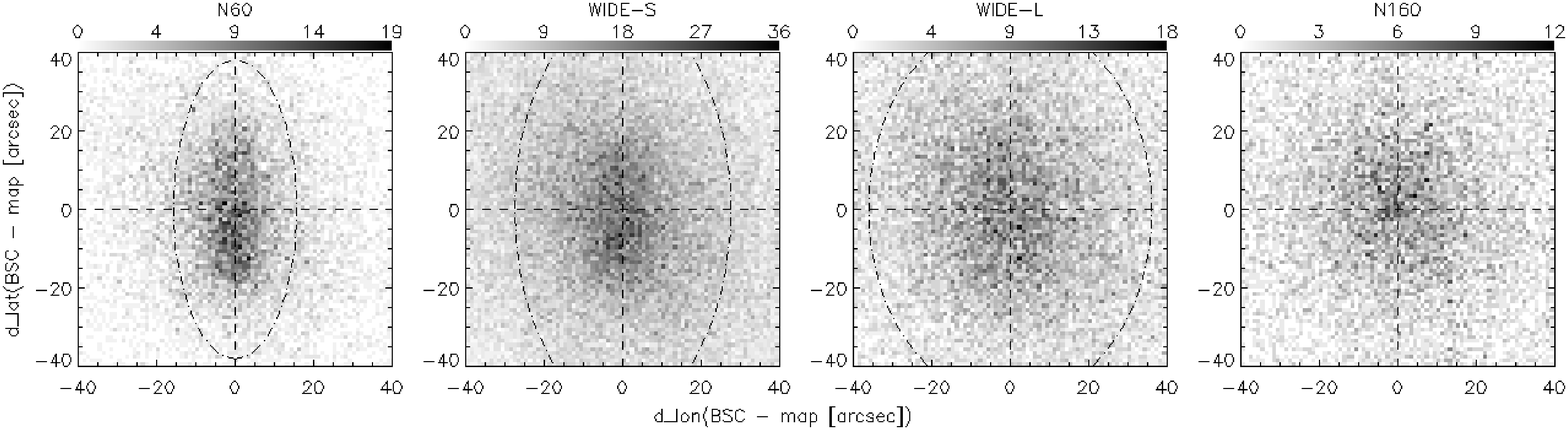}
\end{center}
\caption{
Positional differences of the point sources between the AFASS maps and the FIS BSC for the Normal (top rows) and CDS (bottom rows) modes.
The dash-dotted ovals represent the FWHM of the PSF, tabulated in Table \ref{tbl:psf}.
The differences are estimated on the Ecliptic coordinates, and thus, the x- and y-axis correspond to the cross-scan and in-scan directions of AKARI, respectively.
}
\label{fig:map_pos}
\end{figure*}

We checked the confidence of the data by comparing with the FIS BSC.
Here, we used sources whose quality flags (FQUALxx) are 3 in the BSC, indicating the sources are reliable.
We performed centroid determination for each BSC source using the GCNTRD procedure in the Interactive Data Language (IDL) Astronomy User's Library \citep{landsman1993} to check whether or not the source is detected in the AFASS map.
We rejected those sources whose positional differences between the BSC and the AFASS maps were larger than 48$''$, which was used to extract a unique source in the BSC.
As a result, we have detection rates of $\sim$50~\% for all the 4 bands.
These relatively low detection rates were caused by the large size of the PSFs of the AFASS maps.
Since most of the undetected sources are located around extended objects such as molecular clouds, planetary nebulae and galaxies, it is hard to determine the source centroid in such a region (see Figure~\ref{fig:map_ex}).
We note that the data reduction processes of the BSC were highly optimized for extracting point sources.
Figure~\ref{fig:map_pos} shows the positional accuracy of the AFASS maps.
For the N60 and WIDE-S bands, the accuracy becomes worse along the y-axis, i.e., the scan direction, caused by the elongated shape of the PSFs along the scan direction.

\begin{figure*}
\begin{center}
\includegraphics[width=16cm]{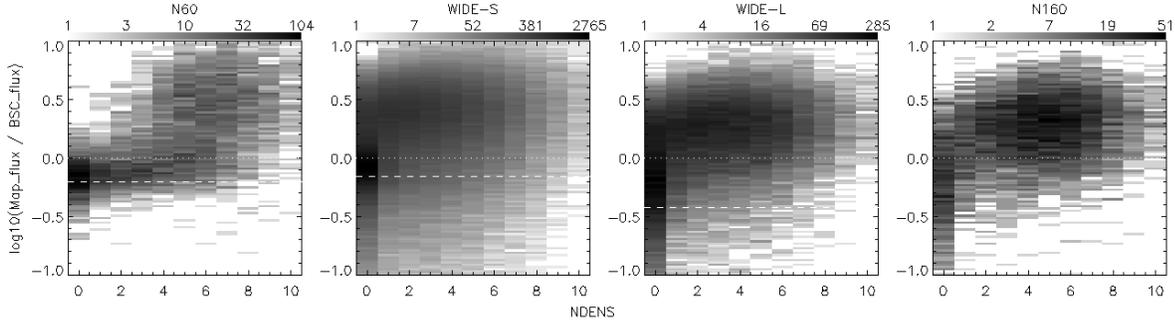}
\end{center}
\caption{
Flux density ratios between the AFASS map and the FIS BSC as a function of the NDENS number.
The dotted lines show the ratio of unity, and the broken lines indicate the ratios derived from the PSF analysis (0.627, 0.696, and 0.381 for the N60, WIDE-S, and WIDE-L bands, respectively).
}
\label{fig:map_bsc_flux_ndens}
\end{figure*}

We also performed aperture photometry for the detected sources using an aperture size of 90$''$ with a sky region of 120--300$''$ in radius.
Since the aperture size was very large, we only considered the source observed in the Normal mode to avoid contaminations from surroundings.
The CDS mode was used near the Galactic plane, and thus, the source density was high.
Figure~\ref{fig:map_bsc_flux_ndens} shows the ratio of the measured to BSC flux densities as a function of NDENS, which is the number of nearby point sources within 5$'$ from each detected source in the BSC.
The ratio is found to become large at NDENS$\ge$1, because of serious contamination caused by the large aperture size.

\begin{figure*}
\begin{center}
\includegraphics[width=16cm]{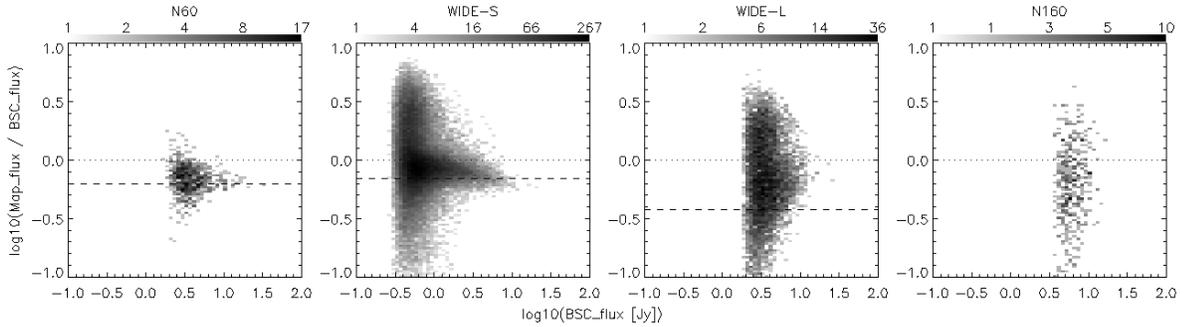}
\end{center}
\caption{
Flux density ratios of the NDENS=0 sources between the AFASS map and the FIS BSC as a function of BSC flux density.
The dotted and broken lines have the same meaning as in Figure~\ref{fig:map_bsc_flux_ndens}.
}
\label{fig:map_bsc_flux}
\end{figure*}

We also show the flux density ratio as a function of the BSC flux density for the sources with NDENS=0 in Figure~~\ref{fig:map_bsc_flux}.
There is no significant correlation with the source flux density, which is consistent with \citet{arimatsu2014}.
The large scatters in flux density ratio are due to the difference in photometry method, i.e., the BSC flux densities are determined by the PSF-fitting method.

\begin{table}
\begin{center}
\caption{Number of the point sources detected in the AFASS maps.}
\label{tbl:bsc_rate}
\begin{tabular}{crrrr}
\hline
& N60 & WIDE-S & WIDE-L & N160\\
\hline
\multicolumn{5}{l}{Normal mode}\\
BSC   & 10799 & 250804 & 66276 & 14722\\
AFASS &  8683 & 166831 & 39415 &  9353\\
\hline
\multicolumn{5}{l}{CDS mode$^{\dagger}$}\\
BSC   & 17980 & 122749 & 52983 & 22135\\
AFASS & 12774 &  50683 & 23139 & 11547\\
\hline
\end{tabular}
\end{center}
($^\dagger$Sources which were took in the CDS mode are indicated by the first bit of the FLAGSxx flag in the BSC.)
\end{table}

\section{Summary}
This paper presents an initial analysis of the imaging performance of the AKARI Far-infrared all sky survey.
We have performed the absolute calibration of the AFASS image by comparing with the COBE/DIRBE data, and the standard deviations with respect to the DIRBE-expected intensities are less than 10\% for intensities above 10, 3, 25, and 26~MJy~sr$^{-1}$ at the N60, WIDE-S, WIDE-L and N160 bands, respectively.
The accuracy at the N60 band was reduced by the unreliability of the zodiacal light model, and those at the two longer wavelengths were due to the detector performances.
The AFASS image is improved in spatial resolution and wavelength coverage, compared with the COBE/DIRBE and IRAS surveys.
The positional accuracies are in good agreement with the PSFs.
The characteristics of point sources in the image were also checked, and are consistent with \citet{arimatsu2014}.

The AFASS data will be useful for studies of the Solar system, the Galaxy and nearby galaxies, in terms of both the spatial distribution of dust and their spectral energy distributions.
It is also helpful for future/on-going infrared/submm missions such as ALMA, JWST, and SPICA.

This research is based on observations with AKARI, a JAXA project with the participation of ESA.
This work has been supported by JSPS KAKENHI Grant Number 19204020, 21111005, 25247016, and 25400220.


\end{document}